\documentclass[showpacs,preprintnumbers,amsmath]{revtex4}
\usepackage{amssymb}
\usepackage{graphicx}
\usepackage{epsfig}
\usepackage{bm}

\makeatletter \leftline{\epsfbox{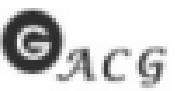}}
 \leftline
{\footnotesize{\textbf{\textsf{GACG-04/14}}}}

\newcommand{\be}{\begin{equation}}
\newcommand{\en}{\end{equation}}
\newcommand{\bea}{\begin{eqnarray}}
\newcommand{\ena}{\end{eqnarray}}

\begin{document}

\title{Closed inflationary universe models in Braneworld Cosmology}
\author{Sergio del Campo$^{1}$, Ram\'on Herrera$^{2}$ and Joel Saavedra$^{1}$}

\address{$^{1}$ Instituto de F\'{\i}sica, Pontificia Universidad Cat\'olica de
Valpara\'{\i}so, Casilla 4950, Valpara\'{\i}so.\\
$^{2}$ Departamento de Ciencias F\'{\i}sicas, Universidad Nacional
Andr\'es Bello, Sazie 2320, Santiago, Chile.}

\date{\today}

\begin{abstract}

In this article we study closed inflationary universe models
proposed by Linde  in a brane world cosmological context. In this
scenario we determine and characterize the existence a of closed
universe, in presence of one self-interacting scalar field with an
inflationary stage. Our results are compared to those found in
General Relativity.
\end{abstract}
\maketitle

\affiliation{\ Instituto de F\'\i sica, Pontificia Universidad Cat\'olica de
Valpara\'\i so, Av. Brasil 2950, Casilla 4059, Valpara\'\i so, Chile.}



\section{\label{sec:level1} Introduction}
From  the super-string or M-theory point of view, our universes
must have dimensionality $N+1$ (where $N$ represent spatial
dimension) greater than four \cite{Green:sg}, in order to cancel
the anomaly in the Type I superstring. Recent developments have
show that standard particles or fields, (described by an open
string) are confined to a $N+1$ manifold (called $N$-brane ),
embedded in a higher dimensional space-time. In this sense the
most popular scenarios are those proposed by Randall and Sundrum
\cite{Randall:1999vf,Randall:1999ee}, where the gravitational
field (described by a closed string) can propagate through the
bulk dimensions. In particular, the cosmological Randall-Sundrum
type II scenario \cite{Randall:1999ee} has received great
attention in the last few years \cite{branecosmology}. This
alternative to Einstein's General Relativity (GR) cosmological
models are called Brane World Cosmologies (BW).

The most spectacular consequence of this scenario is the
modification of the Friedmann equation. In particular, when a five
dimensional model is considered, the matter described by a scalar
field is confined to a four dimensional Brane and  gravity
propagate in the bulk. These kinds of models can be obtained from
a higher superstring theory.  In fact, the strongly coupled
$E_{8}\times E_{8}$ heterotic string theory can be identified as
the eleven dimensional M-theory compactified in an orbifold. The
extra six dimensions on the brane are compactified at a very small
scale ~\cite{witten}. The cosmological solutions in five
dimensions of Horava-Witten theory are described in Refs.
\cite{Lukas:1998yy} and \cite{Lukas:1998qs}. For a comprehensible
review of BW cosmology, see Refs.\cite{lecturer}. Specifically,
consequences of a chaotic inflationary universe scenario in a BW
model was described in Ref.~\cite{maartens}, where it was found
that the slow-roll approximation is enhanced by the modification
of the Friedmann equation.

Recent Observations from the Wilkinson Microwave Anisotropy Probe
(WMAP)~\cite{wmap} combined with the accurate measurement of the
first acoustic Doppler peak of Cosmic Microwave Background
(CMB)~\cite{Ostriker,Bernardis,benoit} are consistent with our
universe having a total energy density that is very close to its
critical value, where the total density parameter has the value
$\Omega =1.02\pm 0.04$. Most people interpret this value as
corresponding to a flat universe, which is consistent with the
standard inflationary prediction~\cite{Guth}. But, according to
this value, we might take the alternative point of view of having
a marginally open ~\cite{ellis-k} or closed universe
~\cite{Lindeclosed} with an inflationary period of expansion at
early time. Therefore, it may be interesting to consider
inflationary universe models in which the spatial curvature is
taken into account. In fact, it is interesting to check if the
flatness in the curvature, as well as in the spectrum, are indeed
reliable and robust predictions of inflation~\cite{Lindeclosed}.
In this sense, the possibility of having inflationary universe
scenarios with negative curvature have been study in Refs.
\cite{linde,extendedramon,braneramon} in the context of GR,
Jordan-Brans-Dicke (JBD) theory and BW, respectively. On other
hand, in the case of $\Omega>1$, the possibilities  of having an
inflationary model, has been considered in Refs.
~\cite{Lindeclosed}, \cite{White} and \cite{Ellis} in Einstein's
GR, and in JBD \cite{extendedclosed} theory. Particularly, the
case with positive curvature has been marginally indicated by the
WMAP recent observations~\cite {Efstathiou:2003hk}.

Therefore, is interesting to study the possibilities of having a
closed inflationary universe from a string cosmological model. The
purpose of the present paper is to study closed inflationary
universe models in the spirit of Linde's work~\cite{Lindeclosed},
where the scalar field is confined to the four dimensional Brane.

The paper is organized as follows: In Sec. II we present the
cosmological equations in brane world cosmology. In Sec. III we
determine the characteristic of a closed inflationary universe
model with a constant potential. In Sec. IV we determine the
characteristic of a closed inflationary scenario with chaotic
potentials. We also, determine the corresponding density
perturbations for our models. In all the cases, our results are
compared to those analogous obtained from Einstein's theory of
gravity.

\section{\label{sec:level2}The cosmological equations in Brane
World Cosmology}

Brane world scenarios inspired by string theory have acquired much
attention in cosmology. In this sense Shiromizu et al.\cite{maeda}
have found that the four-dimensional Einstein equations induced on
the brane and open an interesting scenario to study cosmological
consequence from this model. The Einstein equations on to the
brane can be written as \cite{maeda}
\begin{equation}
G_{\mu \nu }=-\Lambda _{4}g_{\mu \nu }+(\frac{8\pi
}{M_{4}^{2}})T_{\mu \nu }+(\frac{8\pi }{M_{5}^{2}})S_{\mu \nu
}-\mathcal{E}_{\mu \nu }, \label{inducedeq}
\end{equation}
where $T^{\mu \nu }$ is the stress energy-momentum tensor of the
matter in the brane, $S_{\mu \nu }$ is the local correction to
standard Einstein equations due to the extrinsic curvature and
$\mathcal{E}_{\mu \nu }$ are the nonlocal effect corrections from
a free gravitational field, which arise from the projection of the
bulk Weyl tensor. An extended version of Birkhoff's theorem tells
us that if the bulk spacetime is anti-de Sitter (AdS), then
$\mathcal{E}_{\mu \nu }=0$ \cite{lecturer}\cite{Bowcock:2000cq}.
Note that, in this model the matter is confined in the brane and
the gravity can be propagated to the extra dimensions. If we
assume that the matter in the brane is described by a perfect
fluid and considering a four dimensional
Friedmann-Robertson-Walker metric described by
\begin{equation}
\displaystyle d{s}^{2}\,=\,d{t}^{2}\,-\,a(t)^{2}\,\,\,d\Omega _{k}^{2}\,\,,
\label{met}
\end{equation}
where $a(t)$ is the scale factor, $t$ represents the cosmic time
and $d\Omega _{k}^{2}$ is the spatial line element corresponding
to the hypersurfaces of homogeneity, which could represent a
tree-sphere, a tree-plane or a tree-hyperboloid, with values
$k=1,0,-1$, respectively. In the following we will restrict
ourselves to the case $k=1$ only.

When metric~(\ref{met}), with $k=1$, is introduced into
Eq~(\ref{inducedeq}), we obtain the following field equations
\begin{equation}
\displaystyle \,H^2\equiv\left( \frac{\dot{a}}{a}\right)
^{2}=-\frac{1}{a^2}+\frac{8\pi }{3M_{4}^{2}}\,\rho\,
\left(1+\frac{\rho }{2\sigma }\right),  \label{ec2}
\end{equation}
\begin{equation}
\displaystyle \ddot{a}=\frac{8\pi }{3M_{4}^{2}}a\left( -\dot{\phi}%
^{2}+V(\phi )-\frac{1}{8\sigma }(5\dot{\phi}^{2}-2V(\phi ))(\dot{\phi}%
^{2}+2V(\phi ))\right) ,  \label{ec3}
\end{equation}
and for the scalar field
\begin{equation}
\displaystyle \ddot{\phi}\,\,+3\,\frac{\dot{a}}{a}\,\dot{\phi}\,+\frac{dV}{%
d\phi }=0\,\,,  \label{ec1}
\end{equation}
where the dot denotes time derivatives, $H=(\dot{a}/a$) is the
Hubble expansion rate, $\rho =\frac{1}{2}\dot{\phi}^{2}+V(\phi )$
is the energy density of the scalar field and $\sigma$ represent
the brane tension. From now on we will use units where $c=\hslash
=M_{p}=G^{-1/2}=1.$ Note that this
set of equation reduces to the set of Einstein`s field Eqs., in the limits $%
\sigma \longrightarrow $ $\infty $.

\section{\label{sec:level2}Closed inflationary brane world universe with $V=Const.$}
In the spirit of Linde's work \cite{Lindeclosed}, we study a
closed inflationary universe in the context of brane world
cosmology. Firstly let us consider a toy model with the following
step-like effective potential: $V(\phi)=$ 0 at $\phi<$ 0;
$V(\phi)=V=Const.$ at 0$<\phi<\phi_0$ (where $\phi_{0}$ is the
initial value of the inflaton field). Following Linde, we will
also assume that the effective potential sharply rises to
infinitely large values in a small vicinity of $\phi=\phi_0$. This
potential is inspired by supergravity theories \cite{F-Hybrid}.

We consider that the birth of an inflating closed brane world
universe  can be created  ''from nothing'', where this state is
described by $\dot{a}=0$, $\dot{\phi}=0$, $\phi \geq \phi _{0}$,
and the energy density $V_{0}\geq $ $V$, where $V_{0}=V^{*}-\Delta
\,V$ with $V^{*}=\frac{3\,V}{2}$, and $\Delta\,V$ is a small
quantity. Then, the field $\phi$ instantly falls down from the
heights of the potential sharp growth to the plateau ($V(\phi)=V$)
and the potential energy density becomes converted into the
kinetic energy density,
$\frac{\dot{\phi}^2}{2}=\frac{V}{2}-\Delta\,V$.  This produces
that $\dot{\phi}$ changes from a zero value to a negative constant
value. Thus the velocity of the field $\phi$ is given now by
$\dot{\phi}=-\sqrt{V-2\Delta\,V}$. Since this happens in an early
time of the birth of the universe, then one still has $\dot{a}=0$,
and $\dot{\phi}=-\sqrt{V-2\Delta\,V}$ at that times. Then, these
values could be considered as initial conditions when solving the
set of Eqs.(\ref{ec2})-(\ref{ec1}), for $V(\phi)$=$V$=$const.$ in
the interval $0<\phi<\phi_0$.  Now, from the equation (\ref{ec2})
we see that before and after the field instantly falls down to the
plateau, we should necessarily have $\dot{a}=0$ and
$\dot{\phi_0}=-\sqrt{V-2\Delta\,V}$.

Note that, in the regimen where $V=Const.$, the solution of the
scalar field equation (\ref{ec1}), are given by
\begin{equation}
\dot{\phi}(t)=\dot{\phi}_{0}\left( \frac{a_{0}}{a(t)}\right) ^{3}.
\label{campoesc}
\end{equation}
  Due to this, the evolution of the universe rapidly falls into an exponential regimen (inflationary
stages) where the scalar factor becomes $a\sim\,e^{H\,t}$, and
where the Hubble parameter for the brane world reads as follows

\be
H=\sqrt{\frac{8\pi\,}{3}\,V\left[1+\frac{V}{2\sigma}\right]}\label{Hubble}.
\en

Now in the early time, \textit{i.e.} before he inflationary stage
takes place, the resulting equation for the scale factor is
\begin{equation}
\displaystyle \ddot{a}\,=\,\frac{16\pi \,a\,V\,\beta (t)}{3},
\label{ecbeta}
\end{equation}

where we have introduced a small time-dependent function defined
by
$$
\beta (t)=\frac{1}{2V}[V-\dot{\phi}^{2}-\frac{1}{8\sigma }(5\dot{\phi}%
^{2}-2V)(\dot{\phi}^{2}+2V)]
$$
\begin{equation}
=\frac{1}{V}\left[\Delta\,V-\frac{1}{16\sigma}\left(
9V^2-36V\Delta\,V+20\Delta\,V^2\right) \right] \ll 1.
\end{equation}

 We should note that the field equations present a very interesting
solution. First the particular case

\be
\Delta\,V=\frac{1}{10}\left[9V+4\sigma-2\sqrt{9V^2+18V\sigma+4\sigma^2}\right]\equiv\Delta\,V_{static}\label{deltaV},
\en implies $\beta(t)=0$ and from Eq.(\ref{ecbeta}), we see that
the acceleration of the scale  factor is $\ddot{a}=0$. Since
initially $\dot{a}=0$, the universe remains static and the scalar
field $\phi$ moves with the constant speed $\dot{\phi}=\sqrt{V}$.
Secondly, when considering the case $\Delta\,V<
\Delta\,V_{static}$, we get $\beta<0$, and the acceleration of the
scale factor is $\ddot{a}<0$. This corresponds to $3\dot{a}/a<0$
and the universe collapses. Third in the case when
$\Delta\,V>\Delta\,V_{static}$ and one has
$\beta>0\Rightarrow\,\ddot{a}>0$ and $3\dot{a}/a>0$ \textit{i.e.}
the universe enters into an inflationary stage. Note that, as
$\sigma$ goes to infinity, we recover the standard GR results
($\Delta\, V\gtreqqless 0 \Longrightarrow \beta(t)\gtreqqless0$).

 Here, we would like to make a simple analysis of the
solutions of Eq.(\ref {ecbeta}) for $\beta (0)\equiv \beta _{0}\ll
$ 1, in which
\begin{equation}
\beta _{0}=\frac{1}{2V(\phi _{0})}[V((\phi _{0}))-\dot{\phi}^{2}_{0}-\frac{1}{8\sigma }(5\dot{\phi}%
_{0}^{2}-2V(\phi _{0}))(\dot{\phi}_{0}^{2}+2V(\phi _{0}))] .
\label{betacero}
\end{equation}
After $a(t)$ grows  and the inflation phases  settled up, the
inflaton scalar field $\phi $ gradually stops moving. From Eq.
(\ref{campoesc}) together with $a\sim\,e^{H\,t}$, where $H$ is
given by Eq.({\ref{Hubble}}) we have
\begin{equation}
\hspace{-0.1cm}\Delta \phi _{\inf }=\frac{\dot{\phi}_{0}}{3\,H
}\approx -\frac{1}{2}\sqrt{\frac{1}{6\pi }}\frac{1}{(1+V/2\sigma
)^{1/2}}.
\end{equation}
In the limit $\sigma \longrightarrow $ $\infty $, we obtain
$\Delta \phi _{\inf }\approx -1/(2\sqrt{6\pi })$, which coincides
with the result obtained in GR \cite{Lindeclosed}.

When the process starts, $a\approx \,a_{0}$ and $\beta (t)\approx
\beta _{0}$, and Eq. (\ref{ecbeta}) takes the form
\begin{equation}
\displaystyle \ddot{a}\,=\,\frac{16\pi \,a_{0}\,V\,\beta (0)}{3},
\label{ecbetao}
\end{equation}

and hence for small $t$ the solution of Eq. (\ref{ecbetao}), i.e.
the scalar factor $a(t)$, is given by
\begin{equation}
a(t)=a_{0}\left( 1+\frac{8\pi \beta _{0}}{3\,}Vt^{2}\,\right) .
\end{equation}
From Eq. (\ref{campoesc}) we find that at a time interval where
$\beta$ becomes twice as large as $\beta_{0}$, $\Delta \,t_{1}$ is
given by
\begin{equation}
\Delta \,t_{1}\approx \frac{1}{2\sqrt{2\pi V(1+\frac{9}{4\sigma }V)}}\,\,,
\end{equation}
where we have neglected quadratic terms $\beta _{0}^{2}$. This
approximation is justified since the following condition is
satisfied for our model
\begin{equation}
\frac{62\pi ^{2}}{9}\frac{V^{3}\beta _{0}}{\sigma }<<1.
\label{condition}
\end{equation}
In this approximation, it is found that the inflaton field $\phi $
decreased by the amount
\begin{equation}
\Delta \phi _{1}\sim \,\dot{\phi}(0)\Delta \,t_{1}\approx -\frac{1}{2\sqrt{%
2\pi (1+\frac{9}{4\sigma }V)}}\,\,.
\end{equation}
This process continues, after the time $\Delta \,t_{2}\approx \Delta \,t_{1}$%
, where now the field $\phi$ decreases by the amount $\Delta \phi
_{2}\approx \Delta \phi _{1}$, and consequently the rate of growth
of the scalar factor, $a(t)$ increase. This process finishes when
$\beta (t)\rightarrow (1/2+V/4\sigma )$. Therefore, the beginning
of inflation is determined by the initial value of the inflaton
field given by \be
 \phi _{\inf }\approx \phi
_{0}+\frac{1}{2\sqrt{6\pi (1+\frac{1}{2\sigma }V)}}
+\frac{1}{2\sqrt{2\pi (1+\frac{9}{4\sigma }V)}}\ln (\beta
_{0})\label{inflaton}.
\end{equation}
Note that this expression indicates that our results are very
sensitive to the choice of a particular value of  the rate
$\frac{V}{\sigma} $. In the limit $\sigma \,\longrightarrow
\,\infty $, the above expression reduces to $ \phi _{\inf }\approx
\phi _{0}+0.1+0.15\ln \beta _{0}$, where now $\beta _{0}$ becomes, $\beta _{0}\longrightarrow (1-\dot{\phi}%
_{0}^{2}/V)/2$. Since inflation occurs in the interval $\phi
_{\inf }> 0$ and $\phi=0$, the initial value of the inflaton field
becomes \be
 \phi _{0}>-\frac{1}{2\sqrt{6\pi
(1+\frac{1}{2\sigma }V)}}\,- \frac{1}{2\sqrt{2\pi
(1+\frac{9}{4\sigma }V)}}\ln (\beta _{0}). \en

On the other hand, in order to determined the initial value of the
scalar field ($\phi_{0}$), we need to found the value of
$\beta_{0}$. To perform this task, we study the birth of a closed
brane world universe. From the semiclassical point of view, the
probability of creation of a closed universe from nothing in the
brane world scenarios, is given by \cite{LDC}

\begin{equation}
P\,\sim\,e^{-2|S|}\,=\exp\left(\frac{-\pi}{H^2}
\right)=\exp\left(\frac{-3}{8\,V\left[1+\frac{V}{2\sigma}\right]}
\right).
\end{equation}

The probability of creation of the universe  with an energy
density equal to $V^\ast-\beta_{0}\,V$, under the condition that
its energy density $V$ be smaller that $V^\ast$, becomes
\[
P\sim\,\exp\left(\frac{-3}{8}\left[(V^\ast-\beta_0\,V)^{-1}\left[1+\frac{V^\ast-\beta_0\,V}{2\sigma}\right]^{-1}
-\left(V^\ast\left[1+\frac{V^\ast}{2\sigma}\right]  \right)^{-1}
\right]\right),
\]
\begin{equation}
P\,\sim\,\exp\left(-\frac{\beta_0}{6\,V}\left[\frac{
1+\frac{3\,V}{2\,\sigma}}{\left(1+\frac{3\,V}{4\,\sigma}\right)^{2}}\right]
\right)\label{Prob}.
\end{equation}
This latter expression shows that the quantum process of creation
of an inflationary universe model, is not exponentially suppressed
for
\begin{equation} \beta_0<6\,V\,\left[
1+\frac{3\,V}{2\,\sigma}\right]^{-1}\left[
1+\frac{3\,V}{4\,\sigma}\right]^{2} \label{condibeta},
\end{equation}
which means that the initial value of the inflaton field $\phi$
must be bounded from below, \textit{i.e.}

\be \phi _{0}>-\frac{1}{2\sqrt{6\pi (1+\frac{1}{2\sigma }V)}}\,-
\frac{1}{2\sqrt{2\pi (1+\frac{9}{4\sigma }V)}}\ln
\left(\frac{6\,V}{1+\frac{3\,V}{2\,\sigma}}\left[
1+\frac{3\,V}{4\,\sigma}\right]^{2}\right)\label{campoini}. \en It
is straightforward to check that when
$\sigma\longrightarrow\infty$, the GR limit is obtained.

 In order to find the initial value of the scalar field $\phi_{0}$, we
consider some numeric values of the different parameters.
Specifically, we take two different values of $\sigma$,
$\sigma=10^{-10}$ and $\sigma=10^{-9}$. Since we have used units
where the Planck mass in four dimension is equal to one, then the
Planck mass in five dimension becomes  $M_{5}\leq
10^{-2}$~\cite{maartens} and due to this relation, we arrive
$\sigma \thicksim 10^{-10}$. On the other hand, is chosen
$V\sim\,10^{-11}$. As the value for the effective potential
energy, like in the case of chaotic inflationary models, at the
end of the period of inflation \cite{LPC}.

From Eq.(\ref{condibeta}), we obtain $\beta_0<6.0\cdot10^{-11}$
considering the cases $\sigma=10^{-10}$ and $\sigma=10^{-9}$. The
value $\beta_0$ allows us to fix the initial value of the inflaton
field. Table I resumes our results.

\begin{table}
\caption{\label{tab:table1}Initial value of the inlfaton field
$\phi_0$ and $\left| \Delta \phi _{0}\right| _{\inf }$ for two
different values of $\sigma$.}
\begin{ruledtabular}
\begin{tabular}{llll}
& $\phi _{0}>$ & $\phi _{0}\sim $ & $|\Delta\phi_{\inf}|\sim $
\\  \hline $\sigma =10^{-10}$ & $-0.11-0.18\ln
(6\cdot10^{-11})$ & $4.1M_{P}$ & $0.11$ \\ \hline $\sigma
=10^{-9}$ & $-0.11-0.2\ln (6\cdot10^{-11})$ & $4.5M_{P}$ & $0.11$
\\  \hline
\end{tabular}
\end{ruledtabular}
\end{table}

After the inflation, the field $\phi$ stops moving when it passes
the distance $|\Delta\phi_{\inf}|\simeq$ 0.11. However, this
result is a particular value that depends on the value we assign
to the parameter $\sigma $ and the effective potential $V$.

Note that if the field stops before it reaches $\phi=0$, the
universe expands for ever in an inflationary stage. Note that the
same
problem arrives in  Einstein's General Relativity model where it is found that $\Delta\phi_{\inf}%
\simeq-1/2\sqrt{6\pi}=const.$ \cite{Lindeclosed}.  However, in the
context of Brane-World cosmology the value of $\Delta\phi_{\inf}$
depend on the value we assign to the parameter $\sigma$.
Therefore,  we will see that the problem of the universe inflating
forever disappears and thus the inflaton field can reach the value
$\phi =0$ for some appropriate conditions of the ratio $V/\sigma$
that differing from Einstein's GR theory.

 Numerical solutions for the inflaton field $\phi(t)$ are shown in Fig.1 for
two different values of the $\sigma$ parameter. Note that the
interval from $\phi_0$ to $\phi_{\inf}$ increases when the
parameter $\sigma$ decreases, but its shapes remain practically
unchanged. We should note here that, as long as we decrease the
value of the parameter $\sigma$, the quantity $\phi_0-\phi_{\inf}$
increases and thus permits $|\Delta\phi_{\inf}|$ to reach
$\phi=0$, and the inflaton field  does not show oscillations.
Inflation begins immediately if the field $\phi$ starts its motion
with sufficiently small velocity, in analogy with Einstein's GR
theory. If it starts with large initial velocity $\dot{\phi_0}$,
and the universe does not present the inflationary period at any
stage.

\begin{figure}[th]
\includegraphics[height=4.0in,width=3.5in,angle=0,clip=true]{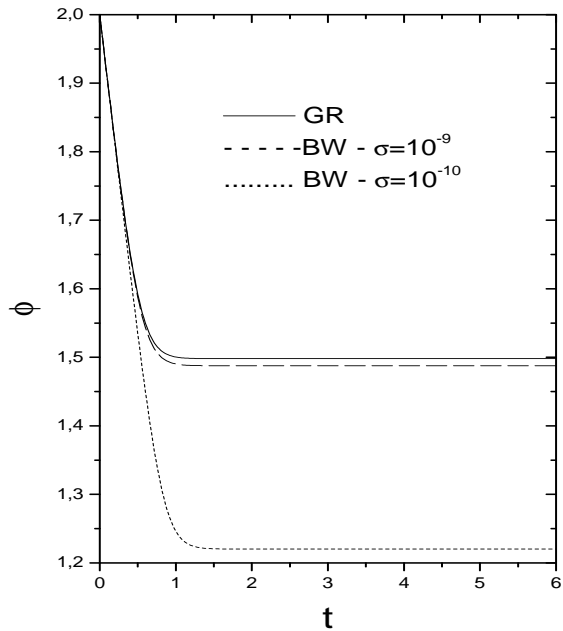}
\caption{Using the model ($V= const$) the inflaton field $\phi$(t)
is plotted as a function of time.  Two different values of the
$\sigma$ parameter are considered, $\sigma=10^{-9}$ and
$\sigma=10^{-10}$. In both cases we have taken the same value of
$\dot{\phi_0}$. GR is displayed on the same plot, but using
Einstein's theory of Relativity.} \label{fig1}.
\end{figure}

\vspace{.5cm}

\section{\label{sec:level2}Brane Chaotic Inflation with $V=\lambda_n\,\phi^n/n$}
The most realistic inflationary universe scenarios are chaotic
models. In this sense, we consider an effective potential given by
$V=\lambda_n\,\phi^n/n$, for $\phi<\phi_0$, which becomes
extremely steep for $\phi > \phi_0$. When the universe is created
at $\phi> \phi_0$, where $V_0>V(\phi_0)=\lambda_n\phi_0^n/n$, the
field immediately falls down to $\phi_0$ and acquires a velocity
given by $\dot{\phi}_0^2/2=V_0-V(\phi_0)$. If the velocity of the
field is small, inflation can start immediately. On the other
hand, if the velocity is large, the universe never inflates.

In order to proceed, we introduce the parameter $\beta_0$, just as
in the previous section. During inflation, the scalar factor is
given by \cite{maartens}
\begin{equation}
\frac{a}{a_o}=\exp\left(-8\pi\,\int_{\phi_o}^{\phi}\,\frac{V[1+V/2\sigma]}{dV/d\phi^{'}}\,d\phi^{'}\right),
\end{equation}
and the corresponding $N$ e-folds, is given by
$$
N=\frac{4\pi\phi_0^2}{n}\;\left[
1+\frac{\lambda_n\,\phi_0^n}{n(n+2)\sigma}\right].
$$

Note that the beginning of inflation is determined by the initial
value of the inflaton field given by Eq. (\ref{inflaton}). We
resume our main results in table II.

\begin{table}
\caption{\label{tab:table2}Values of $\phi_{inf}$ and $\beta_{0}$
for the models with $n=2$ and $n=4$. Parameter values are given by
$\sigma =10^{-10}$ and $V \sim 10^{-11}$}
\begin{ruledtabular}
\begin{tabular}{llll}
& $\phi _{inf}>$ & $\beta_{0}$
\\  \hline $n=2$ & $\phi_{0}-4.2$ & $6\cdot10^{-11}$  \\ \hline $n=4$ & $\phi_{0}-4.2$ & $6\cdot10^{-11}$

 \\
\end{tabular}
\end{ruledtabular}
\end{table}

Notice that, for $\phi_0=10$, inflation starts at
$\phi_{\inf}\sim\,6$, the universe inflates $e^{327}$ times and
becomes flat. The universe inflates $e^{60}$  for $\phi_0=4.4$ and
this leads to $ \Omega=1.1$. Note that in analogy with Einstein's
theory of GR and in order to have the value of $\Omega$ in the
range 1 $\lesssim\Omega<$ 1.1 we require to a fine tuning of the
value of $\phi_0$.

The numerical solution $\phi(t)$ is shown in Fig.2 for two
different models, characterized  for the values of $n$ and
considering the same velocity, $\dot{\phi}_0$ in both cases.
\begin{figure}[th]
\includegraphics[height=5.5in,width=5.0in,angle=0,clip=true]{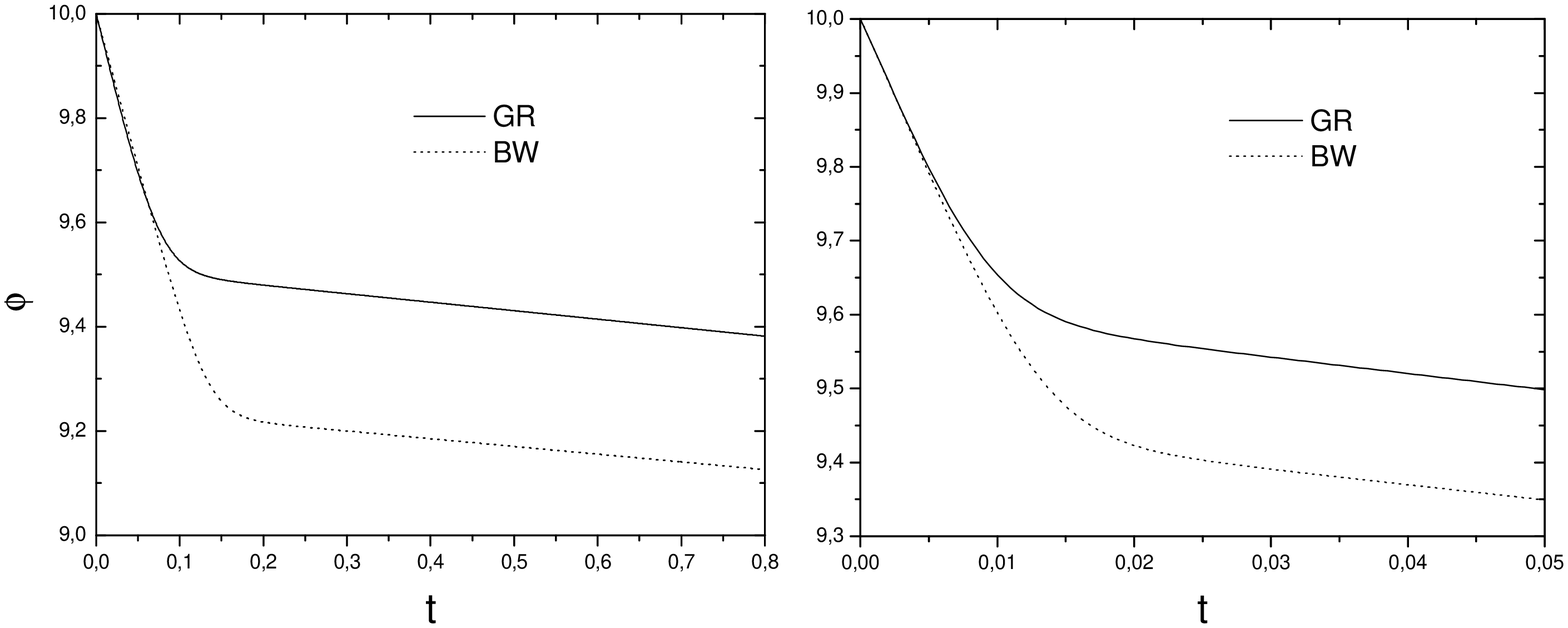}
\vspace{-5.0 cm} \caption{This plot shows the solution for the
inflaton field $\phi(t)$ as a function of cosmological time $t$,
for Einstein's theory of GR and BW theories. The left panel shows
the case $n=2$, and the right one correspond to $n=4$ case. In
both cases we  have taken $\sigma=10^{-10}$. } \label{fig2}.
\end{figure}

One of the main prediction of any inflationary universe models is
the primordial spectrum that arises due to quantum fluctuation of
the inflaton field. Therefore, it is interesting to study the
density perturbation behaviors in brane-world cosmology. We
estimated density perturbations for our models according to
Ref.~\cite{maartens} and thus we may write

\begin{equation}
\displaystyle
\frac{\delta\,\rho}{\rho}\,\approx\,Cte\,\left(\frac{V}{V'}\right)^{\frac{3}{2}}
\left(\,1\,+\,\frac{V}{2\sigma}\right)^{\frac{3}{2}}, \label{ec10}
\end{equation}
where the latter term corresponds to correction due to BW
cosmology for the density perturbations in a flat universe and
$Cte=\frac{24}{5}\sqrt{\frac{8\pi}{3}}$. Certainly, these density
perturbations should be supplemented by several different
contributions for a closed inflationary universe, which alter the
result of $\delta\rho/\rho$ at small $N$. We will postpone this
important matter for a near future. Fig.(3) shows the magnitude of
perturbations in both models, $n=2$ and $n=4$ as a function of $N$
e - folds, for the values $\lambda_2=\frac{9}{4}\cdot\,10^{-12}$,
$\lambda_4=1\cdot\,10^{-14}$ and $\phi_0=10$. Note that
$\delta\rho/\rho$ has a maximum at small $N \simeq 0(7)$, and
presents a small displacement to the right for $\sigma=10^{-10}$.
The maximum is located at $N = 0(8)$, which corresponds to the
scale $\sim$ 10$^{25}$ cm. This latter result is similar to that
obtained in Eisntein's theory of GR. However, the maximum value of
$\delta \rho /\rho $ is bigger in the BW than in for Einstein's
theory of GR.
\begin{figure}[th]
\includegraphics[height=5.0in,width=5.0in,angle=0,clip=true]{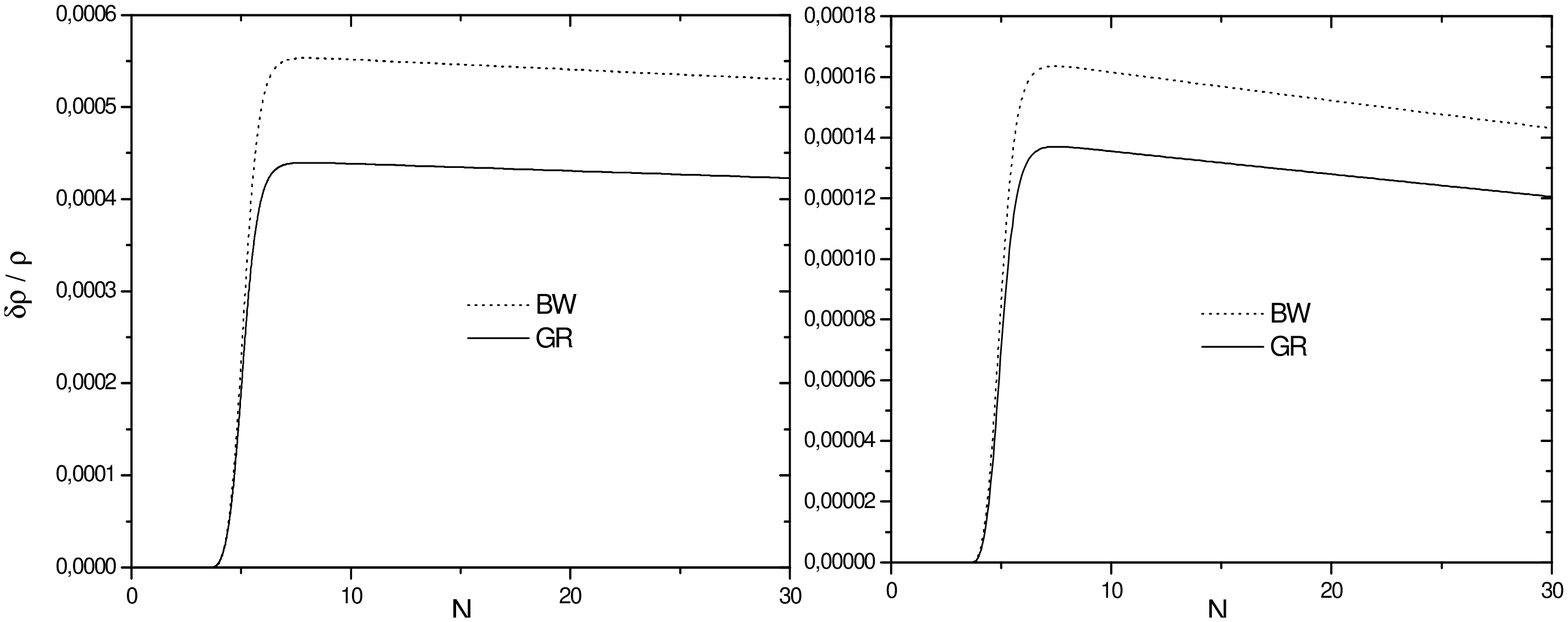}
\vspace{-5.0 cm} \caption{Scalar density perturbation  as a
function of the N e-folds. The left panel shows case $n=2$, and
the right panel shows the case $n=4$. In both cases we  have taken
$\sigma=10^{-10}$. These plots are compared with those obtained by
using Einstein's theory of GR, where $\delta\rho/\rho\approx\,Cte
H^2/|\dot{\phi}|.$} \label{fig3}.
\end{figure}

It is interesting to give an estimation of the tensor spectral
index, $n_T$,  in the Brane-World cosmological model. This index
is given in Ref.\cite{maartens} for a flat universe

\be n_T\simeq\,-\frac{1}{8\pi}\left(\frac{V\,'}{V}\right)^2\,
\left[1+\frac{V}{2\sigma}\right]^{-1}. \en

Solving numerically  for $n=2$ and $n=4$ the field equation
associated with the scalar field $\phi$, we obtain  in Einstein's
theory of GR the following values for the tensor spectral index:
$n_T\simeq\,-15\cdot\,10^{-4}$ which is evaluated for the value of
$N$, where $\delta\rho/\rho$ presents a maximum, i.e, $N\simeq7$
and for $N\simeq6$ $n_T\simeq\,-65\cdot\,10^{-5}$. In BW
cosmological models for $n=2$ we obtain
$n_T\simeq\,-16\cdot\,10^{-4}$ for $N\simeq9$ and for $n=4$,
$n_T\simeq\,-53\cdot\,10^{-5}$ for $N\simeq8$.

\section{Comments and Discussion}
In this article we study a closed inflationary universes with one
self-interacting scalar field in a brane world scenario. For three
different models, corresponding respectively to a constant
potential and two related to a self-interacting scalar potential
given by $V=\lambda_n\,\phi^n/n$ ($n=2$ and $n=4$). In the former
scenarios, we consider a potential with a two regime, one where
the potential is constant and another one where the effective
potential sharply rises to infinity. In the context of Einstein's
theory of GR, this models was study by Linde \cite{Lindeclosed},
who showed that this model is not optimistic due to the constancy
of the potential implying that the universe collapses very soon or
inflates for ever. In our cases, we can fix the graceful exit
problem because in BW cosmology we have an extra ingredient that
is the model dependencies  on the value of the brane tension
$\sigma$, that allowed us to reach the value $\phi$=0, which is
needed to solve this problem.  The problem occurs when $\phi$
reaches the value $\phi=0$, and hence does not show oscillations
for the inflaton field necessary for the reheating process.
However in the latter scenario, this situation disappears in the
chaotic inflationary models, $V(\phi)=\lambda_n\,\phi^n\,/n$.

We have also found that the inclusion of the additional term
($\rho^2$) in the Friedmann's equation change some of the
characteristic of the spectrum of scalar and tensor perturbations.
In this sense, the $\delta\rho/\rho$ graphs presents a small
displacement to the right with respect to $N$, when compared with
that obtained with Einstein`s theory of GR . This would change the
constraint imposed on the value of the parameters that appears in
the scalar potentials $\lambda_n$.  This means that closed
inflationary universe models in a Brane-World theory are less
restricted than those analogous in Einstein`s GR theory.

Note that with the fine-tuning at the level of about one percent,
one can obtain a semi-realistic model of an inflationary universe
with $\Omega>$ 1 as specified by Linde in Ref.\cite{Lindeclosed}.

\begin{acknowledgments}
S.d.C. was supported from COMISION NACIONAL DE CIENCIAS Y
TECNOLOGIA through FONDECYT Grant Nos. 1030469; 1010485 and
1040624. Also, it was partially supported by PUCV  Grant No.
123.764/2004. J.S. was supported from COMISION NACIONAL DE
CIENCIAS Y TECNOLOGIA through FONDECYT Postdoctoral Grant 3030025.
J. S. wish to thank the CECS for its kind hospitality. We thank
Dr. U. Raff for a careful reading of the manuscript.
\end{acknowledgments}

\end{document}